\documentclass[twocolumn, 10pt, journal,letterpaper]{IEEEtran}
\usepackage{url}
\usepackage{graphicx} 
\usepackage{algorithm}
\usepackage[noend]{algpseudocode}
\usepackage{etex} 
\usepackage{tikz}
\usetikzlibrary{shapes,arrows}
\usepackage{pstricks}
\usepackage{pst-node,pst-blur}
\usetikzlibrary{arrows}
\usepackage{amsfonts}
\usepackage{tabularx}
\usepackage{subfigure}
\usepackage{amssymb}
\usepackage{booktabs}
\usepackage{multirow}
\usepackage{steinmetz} 
\usepackage{epsfig}
\usepackage{epstopdf}

\usepackage[noadjust]{cite}

\newtheorem{definition}{Definition}

\usepackage[cmex10]{amsmath} 
\usepackage[cmex10]{mathtools}

\newcommand\norm[1]{\left\lVert#1\right\rVert}

\allowdisplaybreaks  
\makeatletter 
\def\@eqnnum{{\normalsize \normalcolor (\theequation)}} 
\makeatother
\begin{document}
\title{Optimal Least-Squares Estimator and Precoder for Energy Beamforming over IQ-Impaired Channels}
\author{Deepak Mishra,~\IEEEmembership{Member,~IEEE,} and H\aa kan Johansson,~\IEEEmembership{Senior~Member,~IEEE}
	\thanks{D. Mishra and H. Johansson are with the Communication Systems Division of the Department of Electrical Engineering at the Link\"oping University, 581 83 Link\"oping, Sweden (emails: \{deepak.mishra, hakan.johansson\}@liu.se).}
	\thanks{This research work is funded by ELLIIT.}
}

\maketitle
\begin{abstract} 
	Usage of low-cost hardware in large antenna arrays and low-power wireless devices in Internet-of-Things (IoT) has led to the degradation of practical beamforming gains due to the underlying hardware impairments like in-phase-and-quadrature-phase imbalance (IQI). To address this timely concern, we present a new nontrivial closed-form expression for the globally-optimal least-squares estimator (LSE) for the IQI-influenced channel between a multiantenna transmitter and single-antenna IoT device. {Thereafter, to maximize the realistic transmit beamforming gains, a novel precoder design is derived that accounts for the underlying IQI for maximizing received power in both single and multiuser settings.} Lastly, the simulation results, demonstrating a significant $-8$dB improvement in the mean-squared error of the proposed LSE over existing benchmarks, show that the optimal precoder designing is more critical than accurately estimating IQI-impaired channels. Also, the proposed jointly-optimal LSE and beamformer outperforms the existing designs by providing $24\%$ enhancement in the mean signal power received under IQI.
\end{abstract}  

\section{Introduction}\label{sec:intro} 
Massive antenna array technology can help in realizing large beamforming and multiplexing gains~\cite{massive-MIMO}, as desired for the goal of  sustainable ubiquitous Internet-of-Things (IoT) deployment~\cite{SP-Mag-Massive-IoT}. However, due to the  usage of low-cost hardware components, the performance of these sustainable IoT systems is more  prone to suffer from the radio frequency (RF) imperfections~\cite{schenk2008RF} like the in-phase-and-quadrature-phase-imbalance (IQI)~\cite{IQI-2019-Mag}. {Thus, generalized green signal processing techniques are being investigated to combat the adverse effect of hardware impairments~\cite{SPL2-R1-1,SPL2-R1-2,SPL2-R1-3} and the problem of carrier frequency offset (CFO) recovery in frequency-selective IQI~\cite{SPL2-R2-0}.} However, as these impairments adversely influence both channel estimation (CE) and  precoding processes at transmitter (TX), new jointly-optimal estimator and beamformer designs are required.

\subsection{State-of-the-Art}\label{sec:RW} 
In recent times, there have been increasing interests~\cite{IQI-2019-Mag,Virtual-IQI,Eriksson-TC17-IQI,Schober-TWC17-IQI,Access17-IQI,IQI-AF-Stat-MaxD,LMMSE-IQI-Close} on investigating the performance degradation in  energy beamforming (EB) gains of the massive multiple-input-single-output (MISO) systems suffering from IQI. Specifically, as each single-antenna receiver (RX) in multiuser (MU) systems gets wrongly viewed as having a virtual port due to underlying IQI~\cite{Virtual-IQI}, it leads to an inaccurate CE at the multiantenna TX. 
Noting it, sum rate limits in downlink (DL) MU MISO systems under IQI and CE errors were derived in~\cite{Eriksson-TC17-IQI}. In contrast~\cite{Access17-IQI,Schober-TWC17-IQI} were targeted towards the joint CE and IQI compensation in uplink (UL) MISO systems. More recently, performance analysis of  dual-hop statistical channel state information (CSI) assisted cooperative communications was conducted via simulations in~\cite{IQI-AF-Stat-MaxD} to incorporate the effect of IQI. However, these works \cite{Virtual-IQI,Eriksson-TC17-IQI,Schober-TWC17-IQI,Access17-IQI,IQI-AF-Stat-MaxD} only presented linear-minimum-mean-square-error (LMMSE)~\cite{LMMSE-IQI-Close} based CE, that requires strong prior CSI. 

On another front, there are also some works on least-squares (LS) based CE under IQI~\cite{LS-IQI-ICASSP10,ICASSP18,LSE-Pilot-OFDMA}. 
A special structured pilot was used in \cite{LS-IQI-ICASSP10} to obtain LS estimator (LSE) for both actual and IQI-based virtual signal terms. However, these complex pilots are not suited for limited feedback settings involving low-power IoT RX. Therefore, LS and LMMSE estimates using conventional methods were presented in~\cite{ICASSP18} to quantify EB gains during MISO wireless power transfer under joint-TX-RX IQI and CE errors over Rician fading. Lately,  an LSE using additional pilots to exploit the interference among symmetric subcarriers for mitigating effect of IQI was designed in \cite{LSE-Pilot-OFDMA}.

\subsection{Motivation and Scope}\label{sec:motiv} 
All existing works \cite{Virtual-IQI,Eriksson-TC17-IQI,Schober-TWC17-IQI,Access17-IQI,IQI-AF-Stat-MaxD,LMMSE-IQI-Close,LS-IQI-ICASSP10,ICASSP18,LSE-Pilot-OFDMA}, investigating the impact of IQI on CE, considered  the underlying additional virtual signal term as interference, and \textit{simply ignored} the information content in it. Likewise, the current precoder designs for multiantenna TX serving single-antenna RX are based on \textit{suboptimal} maximum ratio transmission (MRT) scheme, ignoring the impairment that signal undergoes due to  IQI. \textit{To the best of our knowledge, the optimal CE and TX precoder designs respectively  minimizing the underlying LS error and maximizing signal power at RX under IQI and CE errors have not been investigated yet.}

Unlike existing works, the proposed \textit{globally-optimal} LSE does not require any prior  CSI. {The adopted  \textit{novel and generic} complex-to-real-domain transformation based methodology to obtain the LSE and precoder in \textit{closed-form} can be extended for investigating designs in MU and multiantenna RX settings.} Lastly, the proposed precoder design holds for \textit{any CE scheme.} 

\subsection{Contribution of This Letter}\label{sec:contrib} 
Our contribution is three-fold. (1) \textit{Global-minimizer of LS error} during CE under TX-RX-IQI is derived in closed-form. (2) \textit{Novel precoder design} is proposed to globally-maximize the nonconvex received signal power over IQI-impaired MISO channels. {Extension of this design to multiuser settings is also discussed.} (3) To validate the nontrivial analysis for different  system parameters, extensive simulations are  conducted, which also \textit{quantify the  achievable EB gains}  over benchmarks. After outlining system model in Section~\ref{sec:model}, these three contributions are discoursed in Sections~\ref{sec:OCE}, \ref{sec:OTB}, and \ref{sec:results}, respectively.

%
\section{System Description}\label{sec:model}    
In this section we present the system model details, followed by the adopted transmission protocol and IQI signal model.
\subsection{Wireless Channel Model and Transmission Protocol} 
We consider DL MISO system comprising of an $N$ antenna source $\mathcal{S}$ and  a single-antenna IoT user $\mathcal{U}$.  Assuming flat quasi-static Rayleigh block fading~\cite[Ch 2.2]{simon2005digital}, the  $\mathcal{U}$-to-$\mathcal{S}$ channel is represented by  $\mathbf{h}\sim\mathbb{C} \mathbb{N}\left(\mathbf{0}_{N\times1},\beta\,\mathbf{I}_N\right)$, where  $\beta$ incorporates the effect of both distance-dependent path loss and shadowing.
%
Transmission protocol  involves estimation of $\mathbf{h}$ from the received IQI-impaired signal at $\mathcal{S}$.  Exploiting  channel reciprocity in the adopted time-division duplex  mode~\cite{CSI-WET-Rician}, we can divide each coherence block of $\tau$ seconds (s) into two phases, namely CE and information transfer (IT). During  CE phase of duration $ \tau_c\le \tau$, $\mathcal{U}$ transmits a pilot signal $\mathrm{s}$ with mean  power $p_c$ and the resulting received baseband signal at $\mathcal{S}$ without any IQI is 
\begin{equation}\label{eq:rxS}
\mathbf{y}=\mathbf{h}\,\mathrm{s} +\mathbf{n},
\end{equation}
where $\mathbf{n}\in\mathbb{C}^{N\times1}$ is  received additive white Gaussian noise (AWGN) vector with  zero mean entries having variance $\sigma_{1}^{2}$.  

\subsection{Adopted Transmission Protocol}\label{sec:prot}  
Our protocol  involves estimation of $\mathbf{h}$ from the received IQI-impaired signal at $\mathcal{S}$.  Here, exploiting  channel reciprocity in the adopted time-division duplex  mode~\cite{CSI-WET-Rician}, we can divide each coherence block into two phases, namely CE and information transfer (IT). During  CE phase of duration $ \tau_c\le \tau$, $\mathcal{U}$ transmits a pilot signal $\mathrm{s}$ with mean  power $p_c$ and the resulting received baseband signal   $\mathbf{y}\in\mathbb{C}^{N\times1}$ at $\mathcal{S}$ without any IQI is 
\begin{equation}\label{eq:rxS}
\mathbf{y}=\mathbf{h}\,\mathrm{s} +\mathbf{n},
\end{equation}  
where $\mathbf{n}\in\mathbb{C}^{N\times1}$ is  received additive white Gaussian noise (AWGN) vector with  zero mean entries having variance $\sigma_{1}^{2}$.  

\subsection{Signal Model for Characterizing IQ Impairments} 
We assume that received baseband signal $\mathbf{y}$ in \eqref{eq:rxS} undergoes the joint-TX-RX-IQI. Therefore, the baseband signal $\mathrm{s}$ at $\mathcal{U}$ gets practically altered to $\mathrm{s}_{\mathrm{T}}$, defined below, due to TX-IQI~\cite{schenk2008RF}
\begin{equation}\label{eq:TXIQI}
\mathrm{s}_{\mathrm{T}}=\mathrm{T}_{\mathcal{U}1}\,\mathrm{s}+\mathrm{T}_{\mathcal{U}2}\,\mathrm{s}^*.
\end{equation}
Here, $\mathrm{T}_{\mathcal{U}1}\triangleq\frac{1+g_{{\mathrm T}_{\mathcal{U}}}\mathrm{e}^{j\phi_{{\mathrm T}_{\mathcal{U}}}}}{2}$ and $\mathrm{T}_{\mathcal{U}2}\triangleq\frac{1-g_{{\mathrm T}_{\mathcal{U}}}\mathrm{e}^{j\phi_{{\mathrm T}_{\mathcal{U}}}}}{2}$, with $g_{{\mathrm T}_{\mathcal{U}}}$ and $\phi_{{\mathrm T}_{\mathcal{U}}}$ respectively denoting TX amplitude and phase mismatch at IoT user $\mathcal{U}$.
Similarly, the baseband signal $\mathbf{y}$ received at $\mathcal{S}$ gets practically impaired due to RX-IQI as~\cite{schenk2008RF}
\begin{equation}\label{eq:RXIQI}
\mathbf{y}_{\mathrm{R}}=\mathbf{R}_{\mathcal{S}1}\,\mathbf{y}+\mathbf{R}_{\mathcal{S}2}\,\mathbf{y}^*,
\end{equation} 
where $i$th diagonal entry of diagonal matrices $\mathbf{R}_{\mathcal{S}1}$ and $\mathbf{R}_{\mathcal{S}2}$ are  $[\mathbf{R}_{\mathcal{S}1}]_i\triangleq\frac{1+g_{{\mathrm R}_{\mathcal{S}i}}\mathrm{e}^{-j\phi_{{\mathrm R}_{\mathcal{S}i}}}}{2}$ and $[\mathbf{R}_{\mathcal{S}2}]_i\triangleq\frac{1-g_{{\mathrm R}_{\mathcal{S}i}}\mathrm{e}^{j\phi_{{\mathrm R}_{\mathcal{S}i}}}}{2}$. Here $g_{{\mathrm R}_{\mathcal{S}i}}$ and $\phi_{{\mathrm R}_{\mathcal{S}i}}$ respectively denote the RX amplitude and phase mismatch at the $i$th antenna of $\mathcal{S}$. 
Finally, combining \eqref{eq:TXIQI} and \eqref{eq:RXIQI} in \eqref{eq:rxS}, the baseband signal $\mathbf{y}_{\mathrm{J}}\in\mathbb{C}^{N\times1}$ as received  at $\mathcal{S}$ during CE phase under joint-TX-RX-IQI is given by 
\begin{equation}\label{eq:JXIQI}
\mathbf{y}_{\mathrm{J}}=\mathbf{h}_{\mathrm A}\,\mathrm{s}+\mathbf{h}_{\mathrm B}\,\mathrm{s}^* +\mathbf{n}_{\mathrm{J}},
\end{equation} 
where $\mathbf{h}_{\mathrm A}\triangleq\mathbf{R}_{\mathcal{S}1}\,\mathbf{h}\,\mathrm{T}_{\mathcal{U}1}+\mathbf{R}_{\mathcal{S}2}\,\mathbf{h}^*\,\mathrm{T}_{\mathcal{U}2}^*$,  $\mathbf{h}_{\mathrm B}\triangleq\mathbf{R}_{\mathcal{S}1}\,\mathbf{h}\,\mathrm{T}_{\mathcal{U}2}+\mathbf{R}_{\mathcal{S}2}\,\mathbf{h}^*\,\mathrm{T}_{\mathcal{U}1}^*$, and $\mathbf{n}_{\mathrm{J}}\triangleq\mathbf{R}_{\mathcal{S}1}\,\mathbf{n} +\mathbf{R}_{\mathcal{S}2}\,\mathbf{n}^*$. We recall that for addressing the demands of low-rate IoT settings using narrow band signals~\cite{Eriksson-TC17-IQI,Access17-IQI}, we have adopted this frequency-independent-IQI model~\cite{schenk2008RF}. Furthermore, as the IQI parameters change very slowly as compared to the channel estimates, we assume their perfect knowledge availability at $\mathcal{S}$~\cite{IQI-2019-Mag,Virtual-IQI,Eriksson-TC17-IQI,Schober-TWC17-IQI,Access17-IQI,schenk2008RF}. Using this practically-motivated assumption, we optimally exploit the information available in the IQI-based virtual signal term $\mathbf{h}_{\mathrm B}\,\mathrm{s}^*$ for designing the LSE and precoder at $\mathcal{S}$. Moreover, using this IQI-knowledge, our proposed solution methodology can also be applied to the frequency-dependent-IQI scenarios. 
 
\section{Optimal Channel Estimation}\label{sec:OCE}    
\subsection{Existing LSE $\widehat{\mathbf{h}}_{\mathrm{A}}$ for IQI-Impaired Channels}\label{sec:existing-LSE}  
Current works~\cite{Virtual-IQI,Eriksson-TC17-IQI,Schober-TWC17-IQI,Access17-IQI,IQI-AF-Stat-MaxD,LMMSE-IQI-Close,LS-IQI-ICASSP10,ICASSP18,LSE-Pilot-OFDMA}  considered  $\mathbf{h}_{\mathrm B} \mathrm{s}^* \!+\mathbf{n}_{\mathrm{J}}$ as the effective noise signal under IQI, and thus, applied conventional pseudo-inverse  method~\cite{kay1993fundamentals} on $\mathbf{y}_{\mathrm{J}}$ in \eqref{eq:JXIQI} with $\norm{\mathrm{s}}^2=p_c\tau_c$, to obtain LSE $\widehat{\mathbf{h}}_{\mathrm{A}}$ for the effective channel $\mathbf{h}_{\mathrm A}$  under IQI, defined below 
\begin{equation}\label{eq:LS}
\widehat{\mathbf{h}}_{\mathrm{A}}=\mathbf{y}_{\mathrm{J}}\;\mathrm{s}^*\left(\mathrm{s}\,\mathrm{s}^*\right)^{-1}
=\mathbf{h}_{\mathrm A} +\widetilde{\mathbf{h}}_{\mathrm{A}},
\end{equation}  
where $\widetilde{\mathbf{h}}_{\mathrm{A}}\triangleq\left(\mathbf{h}_{\mathrm B}\,\mathrm{s}^* +\mathbf{n}_{\mathrm{J}}\right)\mathrm{s}^*\left({p_c\,\tau_c}\right)^{-1}$ is underlying CE error.

\subsection{Proposed LS Approach and Challenges} 
As mentioned in Section~\ref{sec:motiv}, we consider both the terms in $\mathbf{y}_{\mathrm{J}}$, i.e., actual $\mathbf{h}_{\mathrm A}\,\mathrm{s}$ and IQI-based virtual $\mathbf{h}_{\mathrm B}\,\mathrm{s}^*$, containing information on $\mathbf{h}$. Therefore, the proposed optimal LSE $\widehat{\mathbf{h}}$ for $\mathbf{h}$ can be obtained by solving the following  LS problem in $ \mathbf{h}$,
\begin{align}\nonumber
\mathcal{O}_1:\, \underset{ \mathbf{h}}{\text{argmin}}\quad\mathcal{E}\triangleq
\norm{\mathbf{y}_{\mathrm{J}}-\mathbf{h}_{\mathrm A}\,\mathrm{s}-\mathbf{h}_{\mathrm B}\,\mathrm{s}^*}^2.
\end{align}

Although $\mathcal{O}_{\mathrm 1}$ is nonconvex due to the presence of $\mathbf{h}^*$ terms in $\mathbf{h}_{\mathrm A}$ and $\mathbf{h}_{\mathrm B}$, we can characterize all the possible candidates for the optimal solution of $\mathcal{O}_{\mathrm 1}$ by setting derivative of  objective $\mathcal{E}$ to zero and then solve in $\mathbf{h}$. Below, we first simplify  $\mathcal{E}$ as
\begin{align}\label{eq:tr-obj}
\mathcal{E}\!=&\,\mathbf{y}_{\mathrm{J}}^{\rm H}\mathbf{y}_{\mathrm{J}}- \mathbf{y}_{\mathrm{J}}^{\rm H}\mathbf{A}\mathbf{h}\!-\!\mathbf{y}_{\mathrm{J}}^{\rm H} \mathbf{B}\mathbf{h}^*\!+ \mathbf{h}^{\rm H}\mathbf{A}^{\rm H}\mathbf{A}\mathbf{h}-  \mathbf{h}^{\rm H}\mathbf{A}^{\rm H}\mathbf{y}_{\mathrm{J}}+ \nonumber\\
&\; \mathbf{h}^{\rm H}\mathbf{A}^{\rm H}\mathbf{B}\mathbf{h}^*\!-\!\mathbf{h}^{\rm T}\mathbf{B}^{\rm H}\mathbf{y}_{\mathrm{J}}\!+\!\mathbf{h}^{\rm T}\mathbf{B}^{\rm H}\mathbf{A}\mathbf{h}\!+\! \mathbf{h}^{\rm T}\mathbf{B}^{\rm H}\mathbf{B}\mathbf{h}^*, 
\end{align}
where $\mathbf{A}\triangleq\mathbf{R}_{\mathcal{S}1}\,\mathrm{T}_{\mathcal{U}1}\,\mathrm{s}+\mathbf{R}_{\mathcal{S}1}\,\mathrm{T}_{\mathcal{U}2}\,\mathrm{s}^*$ and $\mathbf{B}\triangleq\mathbf{R}_{\mathcal{S}2}\,\mathrm{T}_{\mathcal{U}2}^*\,\mathrm{s}+\mathbf{R}_{\mathcal{S}2}\,\mathrm{T}_{\mathcal{U}1}^*\,\mathrm{s}^*$ are  diagonal matrices. Using  complex-valued differentiation rules~\cite{complex-matrixbook} to the find derivative of \eqref{eq:tr-obj} with respect to $\mathbf{h}$ and setting resultant to zero, gives the following system
\begin{equation}\label{eq:der-tr-obj}
\mathbf{h}^{\rm H}\,\mathbf{Z}_{\rm A} +\mathbf{h}^{\rm T}\,\mathbf{Z}_{\rm B}=\mathbf{y}_{\mathrm{AB}}^{\rm T}, 
\end{equation} 
where $\mathbf{y}_{\mathrm{AB}}\!\!\triangleq\!\mathbf{A}^{\rm T}\mathbf{y}_{\mathrm{J}}^*+\mathbf{B}^{\rm H}\mathbf{y}_{\mathrm{J}}\!\in\!\mathbb{C}^{N\times 1}\!$ and  $\mathbf{Z}_{\rm B}\!\triangleq\!\mathbf{B}^{\rm H}\mathbf{A}\!+\!\mathbf{A}^{\rm T}\mathbf{B}^*\!\in\!\mathbb{C}^{N\times N}$. Here, $\mathbf{Z}_{\rm A}\triangleq\mathbf{A}^{\rm H}\mathbf{A}+ \mathbf{B}^{\rm T}\mathbf{B}^*\!\in\mathbb{R}^{N\times N}$ is a diagonal matrix with $\left[\mathbf{Z}_{\rm A}\right]_i=\left|\left[\mathbf{A}\right]_i\right|^2+\left|\left[\mathbf{B}\right]_i\right|^2,\forall i\in\mathcal{N}=\{1,2,\ldots,N\}.$ 

Though, we have been able to reduce the LS problem $\mathcal{O}_1$ of obtaining optimal LSE $\widehat{\mathbf{h}}$ for  IQI-impaired channels to the nonlinear system of equations \eqref{eq:der-tr-obj} in the complex variable $\mathbf{h},$ solving the latter numerically is computationally-expensive and time-consuming, especially for $N\!\gg\!1$. Therefore, next we propose an \textit{equivalent} complex-to real transformation for \textit{efficiently}  obtaining  unique globally-optimal solution $\widehat{\mathbf{h}}$ of $\mathcal{O}_1$.

\subsection{Closed-Form Expression for Globally-Optimal LSE $\widehat{\mathbf{h}}$}\label{sec:CF-LSE}  
Before deriving $\widehat{\mathbf{h}}$, let us define some key notations below. 
\begin{definition}
	We can define the real composite representations for any complex vector $\mathbf{u}\in\mathbb{C}^{n\times1}$ by $\underline{\mathbf{u}}\in\mathbb{R}^{2n\times 1}$ and for any complex matrix $\mathbf{U}\!\in\!\mathbb{C}^{n_1\times n_2}$ by $\underline{\mathbf{U}}\!\in\!\mathbb{R}^{2n_1\times2n_2}$  as below
	\begin{eqnarray}
	\underline{\mathbf{u}}\triangleq\left[\!\!\begin{array}{cc}
	\mathrm{Re}\{\mathbf{u}\} \\
	\mathrm{Im}\{\mathbf{u}\} \end{array}\!\!\right],\qquad\quad
	\underline{\mathbf{U}}\triangleq\left[\!\!\begin{array}{ccc}
	\mathrm{Re}\{\mathbf{U}\} &\;\, -\mathrm{Im}\{\mathbf{U}\} \\
	\mathrm{Im}\{\mathbf{U}\}&\;\quad
	\mathrm{Re}\{\mathbf{U}\} \end{array}\!\!\right].
	\end{eqnarray} 
\end{definition}
Using above definition, \eqref{eq:der-tr-obj}  can be rewritten in real-domain as
\begin{equation}\label{eq:OCE-0}
\left[\!\!\begin{array}{ccc}
\mathbf{Z}_{\rm A} &\;\,   \mathbf{0}_{N\times N} \\
\mathbf{0}_{N\times N} &\;\, -\mathbf{Z}_{\rm A}  \end{array}\!\!\right]\,\underline{\mathbf{h}}  +\underline{\mathbf{Z}_{\rm B}}\,\underline{\mathbf{h}}  =\underline{\mathbf{y}_{\mathrm{AB}}}. 
\end{equation} 
Recalling $\mathbf{Z}_{\rm A}$ is real while solving \eqref{eq:OCE-0}, the real and imaginary terms of the proposed LSE $\widehat{\mathbf{h}}$ can be analytically expressed  as 
\begin{eqnarray}\label{eq:OCE} 
\left[\!\!\!\begin{array}{cc}
\mathrm{Re}\{\widehat{\mathbf{h}}\}\\ \mathrm{Im}\{\widehat{\mathbf{h}}\}\end{array}    \!\!\!\right]\!\triangleq\!\left[\!\!\!\begin{array}{ccc}
\mathbf{Z}_{\rm A}+\mathrm{Re}\{\mathbf{Z}_{\rm B}\} &  -\mathrm{Im}\{\mathbf{Z}_{\rm B}\} \\
\mathrm{Im}\{\mathbf{Z}_{\rm B}\} 
& -\mathbf{Z}_{\rm A}+\mathrm{Re}\{\mathbf{Z}_{\rm B}\} \end{array}\!\!\!\right]^{-1}\! \underline{\mathbf{y}_{\mathrm{AB}}}\,.\! 
\end{eqnarray} 

\section{Optimal Transmit Beamforming Design}\label{sec:OTB}  
After optimizing LSE using CE phase, now  we optimize the efficiency of IT (phase $2$)  over IQI-impaired DL channel. Metric  to be maximized here by optimally designing precoder $\mathbf{x}\in\mathbb{C}^{N\times 1}$    at $\mathcal{S}$ is the \textit{signal power at $\mathcal{U}$} during IT phase. 
 
\subsection{Conventional Precoder Design}  
With $\mathrm{s_{\mathcal{U}}}$ being unit-energy data symbol, the signal received  at $\mathcal{U}$ due to IT, under perfect CSI and no IQI assumption, is
\begin{equation}\label{eq:DL-ET}
\mathrm{y_{\mathcal{U}}}= \mathbf{h}^{\mathrm T} \,\mathbf{x}\,\mathrm{s_{\mathcal{U}}}+\mathrm{n_{\mathcal{U}}},
\end{equation}
where precoder $\mathbf{x}$ satisfies $\left\lVert\mathbf{x}\right\rVert^2\!\!\le\! p_i,$ with $p_i$ being the transmit power of $\mathcal{S}$ and $\mathrm{n_{\mathcal{U}}}\sim\mathbb{C} \mathbb{N}\left(0,\sigma_2^{2}\right)$ is the received AWGN at  $\mathcal{U}$.  Like in case of CE, the existing  works~\cite{Virtual-IQI,Eriksson-TC17-IQI,Schober-TWC17-IQI,Access17-IQI,IQI-AF-Stat-MaxD,LMMSE-IQI-Close,LS-IQI-ICASSP10,ICASSP18,LSE-Pilot-OFDMA} ignored the virtual term and  designed the precoder as in conventional systems to perform MRT at $\mathcal{S}$ in the DL. Therefore, using the conventional LSE $\widehat{\mathbf{h}}_{\rm A}$ as defined in Section~\ref{sec:existing-LSE}, the benchmark precoder following MRT is given by $\mathbf{x}_{\rm A}\triangleq\frac{\sqrt{p_i}\;\widehat{\mathbf{h}}_{\rm A}^*}{\norm{\widehat{\mathbf{h}}_{\rm A}^*}}.$

\subsection{Maximizing Received Signal Strength under IQI} 
Under joint-TX-RX-IQI, $\mathrm{y_{\mathcal{U}}}$ gets practically impaired to
\begin{align}\label{eq:Rx-signal}
\mathrm{y}_{\mathcal{U}\mathrm{J}}
=\mathrm{R}_{\mathcal{U}1}\,\mathrm{y}_{\mathcal{U}{\rm T}}+\mathrm{R}_{\mathcal{S}2}\,\mathrm{y}_{\mathcal{U}{\rm T}}^*
=\mathbf{a}\,\mathbf{x}+\mathbf{b}\,\mathbf{x}^*+\mathrm{n}_{\mathcal{U}{\rm J}},
\end{align}
where complex vectors $\mathbf{a},\mathbf{b},$ and $\mathrm{n}_{\mathcal{U}{\rm J}}$ are defined below
\begin{subequations}
	\begin{gather} 
	\mathbf{a}\triangleq\left(\mathrm{R}_{\mathcal{U}1}\,\mathbf{h}^{\mathrm T}\,\mathbf{T}_{\mathcal{S}1}+\mathrm{R}_{\mathcal{U}2}\,\mathbf{h}^{\mathrm H}\,\mathbf{T}_{\mathcal{S}2}^*\right)\mathrm{s_{\mathcal{U}}}\in\mathbb{C}^{1\times N},\label{eq:a}\\
	\mathbf{b}\triangleq\left(\mathrm{R}_{\mathcal{U}1}\,\mathbf{h}^{\mathrm T}\,\mathbf{T}_{\mathcal{S}2}+\mathrm{R}_{\mathcal{U}2}\,\mathbf{h}^{\mathrm H}\,\mathbf{T}_{\mathcal{S}1}^*\right)\mathrm{s}_{\mathcal{U}}^*\in\mathbb{C}^{1\times N},\label{eq:b}\\
	\mathrm{n}_{\mathcal{U}{\rm J}}\triangleq\mathrm{R}_{\mathcal{U}1}\,\mathrm{n}_{\mathcal{U}}+\mathrm{R}_{\mathcal{S}2}\,\mathrm{n}_{\mathcal{U}}^*\sim\mathbb{C} \mathbb{N}\left(0,\sigma_{\rm J}^2\right).
	\end{gather}
\end{subequations} 
Here with $g_{{\mathrm R}_{\mathcal{U}}}$ and $\phi_{{\mathrm R}_{\mathcal{U}}}$ respectively denoting  RX amplitude and phase mismatch at  $\mathcal{U}$,   $\mathrm{R}_{\mathcal{U}1}\triangleq\frac{1+g_{{\mathrm R}_{\mathcal{U}}}\mathrm{e}^{-j\phi_{{\mathrm R}_{\mathcal{U}}}}}{2}$ and $\mathrm{R}_{\mathcal{U}2}\triangleq\frac{1-g_{{\mathrm R}_{\mathcal{U}}}\mathrm{e}^{j\phi_{{\mathrm R}_{\mathcal{U}}}}}{2}$.  
Therefore, $\sigma_{\rm J}^2\triangleq\left(\left|\mathrm{R}_{\mathcal{U}1}\right|^2+\left|\mathrm{R}_{\mathcal{U}2}\right|^2\right)\sigma_2^2$.  $\mathrm{y}_{\mathcal{U}{\rm T}}\triangleq\mathbf{h}^{\mathrm T}\left(\mathbf{T}_{\mathcal{S}1}\,\mathbf{x}\,\mathrm{s_{\mathcal{U}}}+\mathbf{T}_{\mathcal{S}2}\,\mathbf{x}^*\mathrm{s_{\mathcal{U}}}^*\right)+\mathrm{n}_{\mathcal{U}}$ is TX-IQI impaired signal, where $\mathbf{T}_{\mathcal{S}1}$ and $\mathbf{T}_{\mathcal{S}2}$ represent  diagonal matrices with $g_{{\mathrm T}_{\mathcal{S}i}}$ and $\phi_{{\mathrm T}_{\mathcal{S}i}}$ in their $i$th diagonal entries $[\mathbf{T}_{\mathcal{S}1}]_i\triangleq \frac{1+g_{{\mathrm T}_{\mathcal{S}i}}\mathrm{e}^{j\phi_{{\mathrm T}_{\mathcal{S}i}}}}{2}$ and $[\mathbf{T}_{\mathcal{S}2}]_i\triangleq \frac{1-g_{{\mathrm T}_{\mathcal{S}i}}\mathrm{e}^{j\phi_{{\mathrm T}_{\mathcal{S}i}}}}{2}$ respectively denoting TX amplitude and phase mismatch at $i$th antenna of  $\mathcal{S}$ during the IT phase. 

Noting that the received signal has two useful terms $\mathbf{a}\,\mathbf{x}$ and $\mathbf{b}\,\mathbf{x}^*$ in \eqref{eq:Rx-signal}, the proposed precoder optimization problem for maximizing the signal power at $\mathcal{U}$ is formulated as below
\begin{equation*}
\mathcal{O}_2\!:\,    \underset{\mathbf{x}}{\text{argmax}}\;\left\lVert\mathbf{a}\,\mathbf{x}+\mathbf{b}\,\mathbf{x}^*\right\rVert^2,\quad\text{subject to}\;\mathrm{(C1)}:\left\lVert\mathbf{x}\right\rVert^2\le p_i.  
\end{equation*}

The challenges here include non-convexity of $\mathcal{O}_2$  and need for fast-converging or closed-form globally-optimal design to obtain  the desired solution $\mathbf{x}_{\rm o}$ in a computationally-efficient manner. Furthermore, this signal power as objective is actually closely-related to other \textit{key metrics} like ergodic capacity and detection error probability~\cite{Opt-R3} because former's higher value   also implies better ergodic capacity  or lower error probability. 

\begin{figure}[!t] 
	\centering\includegraphics[width=3.48in]{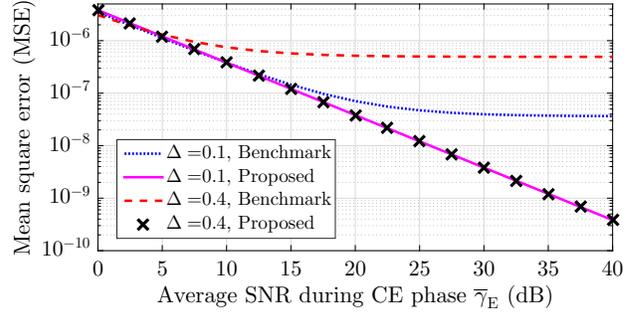} 
	\caption{ Validating  LSE $\widehat{\mathbf{h}}$ against benchmark for different SNR and IQI values.} 
	\label{fig:MSE} 
\end{figure}
\begin{figure*}[!t] 
	\centering\includegraphics[width=6.7in]{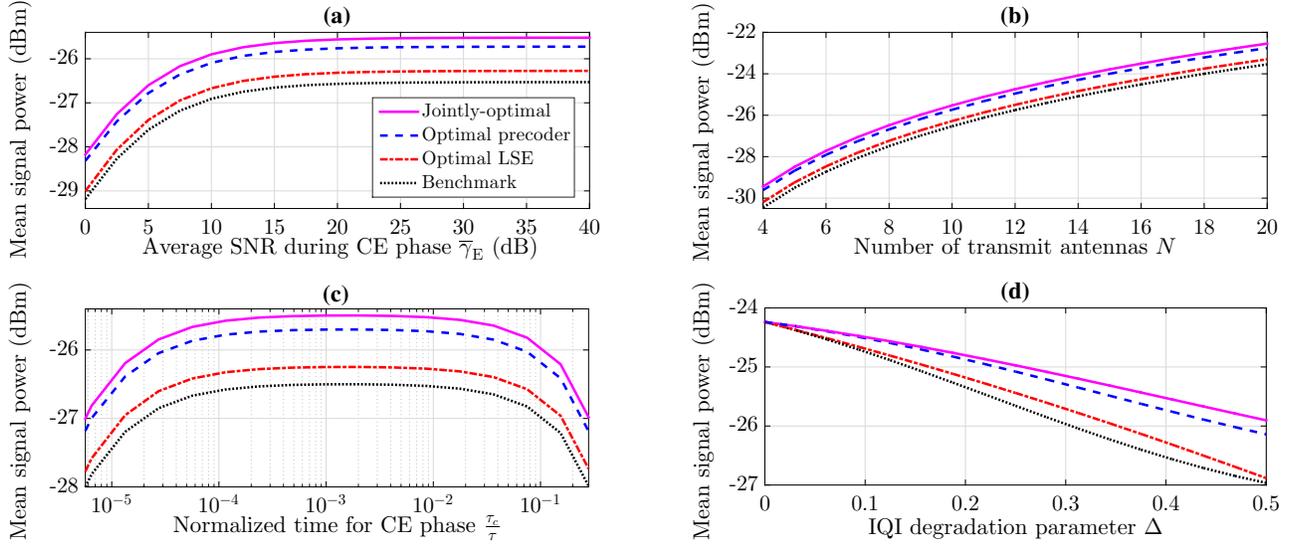}\vspace{-2mm}
	\caption{Comparing relative performance of proposed optimal LSE, precoder, and jointly-optimal designs against benchmark for different $\overline{\gamma}_{\rm E},N,\tau_c,\Delta$ values.}\vspace{-1mm} 
	\label{fig:All}
\end{figure*} 
\subsection{Novel Globally-Optimal Precoder}\label{sec:optP}   
Though $\mathcal{O}_2$ is nonconvex, its globally-optimal solution  can be characterized via  Karush-Kuhn-Tucker (KKT) point~\cite{Baz}. To obtain latter, below we define  Lagrangian function for $\mathcal{O}_2$
\begin{align}
\mathcal{L}\!\triangleq&\left\lVert\mathbf{a}\,\mathbf{x}+\mathbf{b}\,\mathbf{x}^*\right\rVert^2-\ell\,\big(\!\left\lVert\mathbf{x}\right\rVert^2-p_i\big)= \mathbf{x}^{\rm H}\mathbf{a}^{\rm H}\mathbf{a}\,\mathbf{x}+ \mathbf{x}^{\rm H}\mathbf{a}^{\rm H}\mathbf{b}\,\mathbf{x}^*\nonumber\\
&\qquad\;  +\mathbf{x}^{\rm T}\,\mathbf{b}^{\rm H}\mathbf{a}\,\mathbf{x} + \mathbf{x}^{\rm T}\mathbf{b}^{\rm H}\,\mathbf{b}\,\mathbf{x}^*-\ell\,\big(\mathbf{x}^{\rm H}\mathbf{x}-p_i\big),
\end{align} 
where $\ell\ge0$ is the Lagrange multiplier corresponding to $\mathrm{(C1)}$.
\begin{eqnarray}\label{eq:partial}
\frac{\partial\mathcal{L}}{\partial\mathbf{x}}=\mathbf{x}^{\rm H}\mathbf{a}^{\rm H}\mathbf{a}   +\mathbf{x}^{\rm T}\left(\mathbf{b}^{\rm H}\mathbf{a}+\mathbf{a}^{\rm T}\mathbf{b}^*\right)+ \mathbf{x}^{\rm T}\mathbf{b}^{\rm H}\,\mathbf{b}-\ell\, \mathbf{x}^{\rm H}. 
\end{eqnarray}
Setting $\frac{\partial\mathcal{L}}{\partial\mathbf{x}}$ in \eqref{eq:partial} to $\mathbf{0}_{1\times N}$, yields the KKT condition below
\begin{align}\label{eq:KKT1}
\mathbf{x}^{\rm T}\,\mathbf{Z}_{\mathbf a}^*+\mathbf{x}^{\rm H}\,\mathbf{Z}_{\mathbf b}^*=\ell\, \mathbf{x}^{\rm T},
\end{align}
where $\mathbf{Z_a}\triangleq\mathbf{a}^{\rm H}\mathbf{a}+\mathbf{b}^{\rm T}\mathbf{b}^*$ and $\mathbf{Z_b}\triangleq\mathbf{b}^{\rm H}\mathbf{a}+\mathbf{a}^{\rm T}\mathbf{b}^*$. Using the composite real definition from Section~\ref{sec:CF-LSE} in \eqref{eq:KKT1}, we obtain
\begin{eqnarray}\label{eq:Eig}  
\left(\underline{\mathbf{Z}_{\mathbf{a}}}\right)^*\underline{\mathbf{x}}  +\left(\underline{\mathbf{Z}_{\mathbf{b}}}\right)^{\rm H}\left[\!\!\begin{array}{cc}
\;\;\,\mathrm{Re}\{\mathbf{x}\} \\
-\mathrm{Im}\{\mathbf{x}\} \end{array}\!\!\right]= \ell\,\underline{\mathbf{x}},\hspace{2mm}\text{or}\hspace{2mm}\mathbf{Z}_{\mathbf{ab}}\,\underline{\mathbf{x}} =  \ell\,\underline{\mathbf{x}},
\end{eqnarray} 
{where the real square matrix $\mathbf{Z}_{\mathbf{ab}}\in\mathbb{R}^{2N\times2N}$ is defined as}
\begin{eqnarray}\label{eq:Zab} 
\mathbf{Z}_{\mathbf{ab}}\triangleq\left[\!\!\begin{array}{ccc}
\hspace{3mm}\mathrm{Re}\{{\mathbf{Z}_{\mathbf{a}}}\}+\mathrm{Re}\{{\mathbf{Z}_{\mathbf{b}}}\}&\; \mathrm{Im}\{{\mathbf{Z}_{\mathbf{a}}}\}+\mathrm{Im}\{{\mathbf{Z}_{\mathbf{b}}}\}\\
-\mathrm{Im}\{{\mathbf{Z}_{\mathbf{a}}}\}+\mathrm{Im}\{{\mathbf{Z}_{\mathbf{b}}}\}&\; 
\mathrm{Re}\{{\mathbf{Z}_{\mathbf{a}}}\}-\mathrm{Re}\{{\mathbf{Z}_{\mathbf{b}}}\} \end{array}\!\!\right].
\end{eqnarray}   
As \eqref{eq:Eig} possesses an eigenvalue problem form, the solution to \eqref{eq:Eig} in $\underline{\mathbf{x}}$ is given by the principal eigenvector $\mathrm{v}_{\max}\left\lbrace\mathbf{Z}_{\mathbf{ab}}\right\rbrace$ corresponding to the maximum eigenvalue  $\lambda_{\max}\left\lbrace\mathbf{Z}_{\mathbf{ab}}\right\rbrace$ of $\mathbf{Z}_{\mathbf{ab}}$. Therefore, the globally-maximum signal power is attained at the proposed precoder  $\mathbf{x}_{\rm o}\triangleq\mathrm{Re}\{\mathbf{x}_{\rm o}\}+j\,\mathrm{Im}\{\mathbf{x}_{\rm o}\}$, whose real and imaginary parts, obtained via eigen-decomposition are  
\begin{eqnarray}\label{eq:optx}
\underline{\mathbf{x}_{\rm o}}=\left[\begin{array}{cc}
\mathrm{Re}\{\mathbf{x}_{\rm o}\} \\
\mathrm{Im}\{\mathbf{x}_{\rm o}\} \end{array}\right]\triangleq\sqrt{p_i}\;\frac{\mathrm{v}_{\max}\left\lbrace\mathbf{Z}_{\mathbf{ab}}\right\rbrace}{\left\lVert\mathrm{v}_{\max}\left\lbrace\mathbf{Z}_{\mathbf{ab}}\right\rbrace\right\rVert}\in\mathbb{R}^{2N\times1}.
\end{eqnarray}

\subsection{Extending Precoder Design to Multiuser Settings}\label{sec:extn}
For maximizing the sum received power among $K$ single-antenna users, the precoder optimization problem   is given by
\begin{equation*}
\mathcal{O}_3\!:\,    \underset{\mathbf{x}}{\text{argmax}}\;\left\lVert\boldsymbol{\mathcal{A}}\,\mathbf{x}+\boldsymbol{\mathcal{B}}\,\mathbf{x}^*\right\rVert^2,\qquad\text{subject to}\quad\mathrm{(C1)},  
\end{equation*}
where the $K\times N$ matrices $\boldsymbol{\mathcal{A}}$ and $\boldsymbol{\mathcal{B}}$ are respectively obtained from $\mathbf{a}$ and $\mathbf{b}$ in \eqref{eq:a} and  \eqref{eq:b}, but with  $\mathrm{s_{\mathcal{U}}}$ replaced by unit-energy vector $\mathbf{s_{\mathcal{U}}}$, $\mathbf{h}$ replaced with $N\times K$ matrix $\mathbf{H}$ whose $i$th column corresponds to channel gain for $\mathcal{S}$ to $i$th user link, and the $K\times K$ diagonal matrices $\mathbf{R}_{\mathcal{U}1}$ and  $\mathbf{R}_{\mathcal{U}2}$, respectively replacing $\mathrm{R}_{\mathcal{U}1}$ and  $\mathrm{R}_{\mathcal{U}2}$. Here,   $i$th diagonal entries of $\mathbf{R}_{\mathcal{U}1}$ and $\mathbf{R}_{\mathcal{U}2}$ incorporate the  RX amplitude and phase mismatch at $i$th user. So, following Section~\ref{sec:optP}, the optimal precoder for $\mathcal{O}_3$ is given by \eqref{eq:optx}, but with $\boldsymbol{\mathcal{A}}$ and $\boldsymbol{\mathcal{B}}$ respectively replacing $\mathbf{a}$ and $\mathbf{b}$  in $\mathbf{Z}_{\mathbf{ab}}$ definition. The accuracy of this TX design in multiuser setting can be verified from the fact that for no IQI, $\mathbf{Z}_{\mathbf{ab}}$ reduces to $\underline{\mathbf{H}\,\mathbf{H}^{\rm H}}$ with result matching  \cite[Theorem 1]{EBF-TWC14}.\color{black}

\section{Performance Evaluation and Conclusion}\label{sec:results} 
Here we numerically validate the proposed CE analysis and precoder optimization while setting simulation parameters as $N=10$, $\tau=10$ms, $\tau_c=0.01\tau$, $p_i=30$dBm, $p_c=-30$dBm, $\sigma_{1}^{2}=\sigma_{2}^{2}=10^{-17}$ Joule,  and $\beta=\frac{\varpi}{d^{\varrho}}$, where $\varpi=\left(\frac{3\times 10^8}{4\pi f}\right)^2$ is average channel attenuation at unit reference distance with $f=915$MHz  as TX frequency, $d=100$m as $\mathcal{S}$-to-$\mathcal{U}$ distance, and $\varrho=2.5$ as path loss exponent. For the average simulation results, we have used $10^5$ independent channel realizations. 

\subsection{Validation of Proposed LSE under Practical IQI Modelling}\label{sec:sim-valid}   
We start with verifying the quality of proposed LSE $\widehat{\mathbf{h}}$ (cf. \eqref{eq:OCE}) in Fig.~\ref{fig:MSE} against the benchmark   $\widehat{\mathbf{h}}_{\rm A}$ as defined in \eqref{eq:LS}. For IQI incorporation, we adopt the following practical model~\cite{TCOM-HEB} 
\begin{eqnarray}\label{eq:g-p}
g\triangleq1-\Delta_{g}\left(1+\Psi_{g}\right),\qquad\quad\phi\triangleq \Delta_{\phi}\left(1+\Psi_{\phi}\right), 
\end{eqnarray} 
where $g$ and $\phi$ respectively can incorporate any amplitude and phase mismatch, with the constants $\Delta_{g}$ and  $\Delta_{\phi}$ representing the errors due to fixed sources. Whereas,  $\Psi_{g}$ and $\Psi_{\phi}$, respectively  denoting errors due to random sources, are assumed to follow the uniform distribution~\cite{IQI-AF-Stat-MaxD,LMMSE-IQI-Close} over the  interval $\left[-\frac{1}{2}\Phi_{g},\frac{1}{2}\Phi_{g}\right]$ and $\left[-\frac{1}{2}\Phi_{\phi},\frac{1}{2}\Phi_{\phi}\right]$, respectively.  {Since the practical ranges for the constants $\left(\Delta_g,\,\Delta_{\phi},\,\Phi_g,\,\Phi_{\phi}\right)$ corresponding to the means and variances of amplitude and phase errors (in radians) are similar~\cite{TWC-IQI,SPL-HEB}, we set $\Delta_{g}=\Delta_{\phi}=\Phi_{g}=\Phi_{\phi}=\Delta=0.4$, for each of the $8$ IQI parameters.}

Results plotted in Fig.~\ref{fig:MSE} show  the trend in mean square error (MSE)~\cite{massive-MIMO} between the actual channel $\mathbf{h}$ and its LSE (proposed $\widehat{\mathbf{h}}$ and benchmark $\widehat{\mathbf{h}}_{\rm A}$) against increasing average received signal-to-noise-ratio (SNR) $\overline{\gamma}_{\rm E}=\frac{\beta\,p_c\tau_c}{\sigma_1^2}$ at $\mathcal{S}$ during CE phase. The quality of both proposed and existing LSE improve with increasing $\overline{\gamma}_{\rm E}$ because the underlying CE errors reduce for both  considered values of IQI degradation parameter $\Delta$. However, for the benchmark LSE, the error floor region starts at $\overline{\gamma}_{\rm E}=20$dB and $\overline{\gamma}_{\rm E}=30$dB for $\Delta=0.4$ and $\Delta=0.1$, respectively. Whereas, MSE for the proposed globally-optimal LSE for the IQI-impaired channel keeps on  decreasing at the same rate \textit{without} having any error floor. This corroborates the \textit{significantly-higher} practical utility of our proposed LSE $\widehat{\mathbf{h}}$ for the IQI-influenced MISO communications, in terms of our CE design providing about $-3$dB and $-11$dB improvement in MSE over benchmark  for $\Delta=0.1$ and $\Delta=0.4$, respectively.

\subsection{Comparison of Proposed Designs Against Benchmark}\label{sec:comp} 
{Here we compare the mean signal power performance  of the three proposed schemes: (i) \textit{jointly-optimal} LSE $\widehat{\mathbf{h}}$ and precoder $\mathbf{x}_{\rm o}$, (ii) \textit{optimal precoder} $\mathbf{x}_{\rm o}$ with conventional LSE $\widehat{\mathbf{h}}_{\rm A}$,  (iii) \textit{optimal LSE} $\widehat{\mathbf{h}}$ with MRT-based precoder $\mathbf{x}_{\rm A}$, against the \textit{benchmark} having LSE $\widehat{\mathbf{h}}_{\rm A}$ and precoder $\mathbf{x}_{\rm A}$.} Starting with comparison for different  $\overline{\gamma}_{\rm E}$  in Fig.~\ref{fig:All}(a), we notice that jointly-optimal performs the \textit{best}, followed by optimal precoder and proposed LSE. The gaps between the optimal and benchmark designs  increase with  $\overline{\gamma}_{\rm E}$ due to lower errors at higher SNRs.
 
Next in Fig.~\ref{fig:All}(b), we plot the comparison for different array sizes $N$ at $\mathcal{S}$. Here, with $N$ increased from $4$ to $20$, mean signal power at $\mathcal{U}$ gets enhanced by $7$dB for each of the four schemes. However, their relative gap remains \textit{invariant} of $N$.  

Now, shifting focus to CE time $\tau_c$, we shed insights on how to optimally set it. From Fig.~\ref{fig:All}(c), we notice that the relative trend among four schemes is similar, but more importantly, the optimal $\tau_c$ for each scheme is  \textit{practically the  same} ($\approx 10^{-3}\tau$).
   
Next we investigate the impact of increased mismatch $\Delta$ in the amplitude and phase terms modelling the IQI. In particular, by plotting the variation of $\Delta$ from $0$ to $0.5$~\cite{IQI-AF-Stat-MaxD,LMMSE-IQI-Close,TCOM-HEB} in Fig.~\ref{fig:All}(d), we observe that degradation in the mean signal power  performance gets enhanced with increased IQI (i.e., $\Delta$) for each  scheme. However, this performance degradation for jointly-optimal, optimal precoder, optimal LSE, and benchmark schemes when parameter $\Delta$ increases from $0$ (no IQI) to $0.5$ is $-1.6$dB,$-1.9$dB,$-2.7$dB, and $-2.8$dB, respectively.  

\begin{figure}[!t] 
	\centering\includegraphics[width=3.4in]{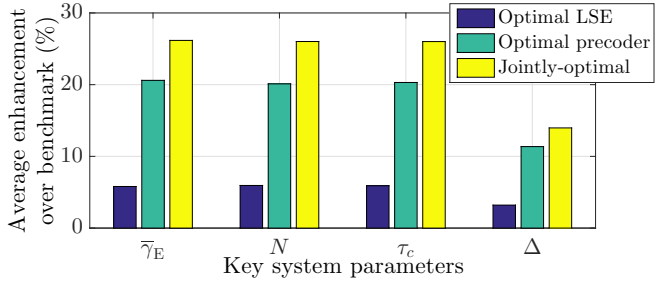}\vspace{-2mm}
	\caption{Average performance gains of our proposed designs over benchmark.}
	\label{fig:comp} 
\end{figure}  

Lastly, in Fig.~\ref{fig:comp}, we have plotted the average performance gains as achieved by the proposed LSE $\widehat{\mathbf{h}}$, precoder $\mathbf{x}_{\rm o}$, and the jointly-optimal design over benchmark for different values of critical parameters $\overline{\gamma}_{\rm E},N,\tau_c,$ and $\Delta$.  {We observe that jointly-optimal design provides an \textit{overall improvement of $24$\%}.} Here, optimal precoder, providing about $18$\% enhancement alone in mean signal power at $\mathcal{U}$, proved to be a \textit{better semi-adaptive scheme} than optimizing LSE, which yields  $6\%$ improvement.


\subsection{Concluding Remarks}\label{sec:conclusion}
This letter exploiting the additional channel gain information in the signal received during IQI-impaired MISO communication, came up with a \textit{novel LSE} that is shown to  reduce the overall MSE in CE by $-8$dB, while totally \textit{removing the error floor}. {To maximize the practical EB gains in both single and multiple user set-ups, we derive new \textit{globally-optimal precoder} in the form of \textit{principal eigenvector} of the matrix composed of IQI parameters and LSE.} Numerical results have shown that  the proposed jointly-optimal LSE and precoder design can provide an overall improvement of $24\%$ over the benchmark. This corroborates the fact that our proposed design is the \textit{way-forward} to maximize practical utility of low-cost hardware in multiantenna transmission supported sustainable IoT systems.


\end{document}